# Coexistance of volatile and non-volatile memristive effects in phase-separated $La_{0.5}Ca_{0.5}MnO_3$-based devices

G.A. Ramírez [1,2], W. Román Acevedo [1,2], M. Rengifo [3,4,5], J. M. Nuñez [2,3,4,5], M.H. Aguirre [3,4,5], J. Briático [6], and Diego Rubi [1,2]

[1] *Departamento de Micro y Nanotecnologías (CNEA),*
*Gral. Paz 1499 (1650), San Martín, Argentina*

[2] *Instituto de Nanociencia y Nanotecnología (INN),*
*CONICET-CNEA, Buenos Aires and Bariloche, Argentina*

[3] *Departamento de Física de Materia Condensada,*
*Universidad de Zaragoza, Pedro Cerbuna 12, 50009 Zaragoza, Spain*

[4] *Laboratorio de Microscopías Avanzadas (LMA),*
*Instituto de Nanociencia de Aragón (INA)-Universidad de Zaragoza,*
*C/Mariano Esquillor s/n. 50018 Zaragoza, Spain*

[5] *Instituto de Ciencias de Materiales de Aragón (ICMA),*
*Universidad de Zaragoza, 50009 Zaragoza, Spain*

[6] *Unité Mixte de Physique CNRS/Thales and*
*Université Paris-Saclay, Palaiseau 91767, France*

In this work, we have investigated the coexistance of volatile and non-volatile memristive effects in epitaxial phase-separated $La_{0.5}Ca_{0.5}MnO_3$ thin films. At low temperatures (50 K), we observed volatile resistive changes arising from self-heating effects in the vicinity of a metal-to-insulator transition. At higher temperatures (140 K and 200 K) we measured a combination of volatile and non-volatile effects arising from the synergy between self-heating effects and ferromagnetic-metallic phase growth induced by an external electrical field. The results reported here add phase separated manganites to the list of materials which can electrically mimic, on the same device, the behavior of both neurons and synapses, a feature that might be useful for the development of neuromorphic computing hardware

The exponential growth of Artificial Intelligence (AI) -including Machine Learning (ML) algorithms-, applied nowadays to tackle an enormous variety of scientific and technological problems, is pushing modern computers -based on the so-called Von-Neumann architecture, where the processing and storage of information take place in physically separated units- to their limits. In addition, this computing paradigm raises severe issues about the ever-growing carbon footprint related to the massive implementation of AI algorithms [1]. This triggers the search of novel computer architectures and information processing strategies, more energy efficient and specifically tailored to address complex problems such as image or speech recognition, which usually require the processing in parallel of large amounts of data structured in high dimension matrices.

Neuromorphic computing -which intends to develop hardware that mimics the structure and information processing mechanisms of biological systems- appears as a promising technology to accomplish the previously mentioned tasks [1, 2]. For that, it is necessary to develop solid-state devices able to electrically replicate the behavior of both neurons and synapses. Neuromorphic functionalities have been reported for memristive [2–4], ferroelectric [5–7], phase change [8–10] and spintronic [11–13] devices, which are capable of modifying their electrical properties through controlled external stimuli. Memristors are metal/insulator/metal micro or nanostructures that allow a reversible change of their electrical resistance upon the application of electrical stress [14, 15]. In many cases, these changes are non-volatile and analog, replicating the behavior of biological synapses [2]; indeed, cross-bar arrays of memristors have been shown as a possible hardware implementation of neural networks, where the network synaptic weights are linked to memristor conductivities [16].

For some memristive systems, the resistance change is volatile -that is, the resistance changes during the application of external stress and relaxes to its original state after the removal of the stimuli-, and are therefore suitable for mimicking different functionalities of biological neurons with different complexity, from integrate-and-fire to chaotic oscilations [17]. Volatile memristive behavior have been mostly reported for devices displaying electrochemical metalization -the volatile effect is related to the spontaneous dilution of Cu or Ag nanofilaments- [18, 19] and also for correlated or Mott insulators such as $VO_2$, $V_2O_3$ or $NbO_2$, where thermal effects control the resistive change in the vicinity of insulator-metal transitions (see Ref. [20] and references therein). Volatile memristive effects have been also found in ferroelectric-based systems, and the resistance relaxations have been related either to polarization loss [21] or oxygen vacancy electromigration driven by the depolarizing field [22]. Volatile memristors might be implemented in applications such as selectors in cross-bar arrays, true random generators, physical unclonable functions and also in reservoir computing (see Ref. [23] and references therein).

Interesting alternative systems to explore the existence of volatile and non-volatile memristive effects are the phase separated manganites. In these systems, phase coexistence usually arises from the competition of ferromagnetic-metallic (FM-M) and antiferromagnetic-insulating (AF-I) phases, presenting similar energies, at the nano or mesoscopic scale [24]. Structural features such as the grain size in polycrystalline samples [25] or epitaxial strain [26] in thin films determine the balance between FM-M and AF-I phases, which, in addition, can be tuned by using a number of external knobs such as temperature [27], magnetic [28] or electrical fields [29]. Non-volatile memristive effects have been reported for devices based on phase separated manganites such as $La_{0.5}Ca_{0.5}MnO_3$ [29, 30], $La_{0.3}Pr_{0.4}Ca_{0.3}MnO_3$ [31], $Nd_{0.5}Sr_{0.5}MnO_3$ [32] or $Sm_{0.5}Ca_{0.25}Sr_{0.25}MnO_3$ [33].

$La_{0.5}Ca_{0.5}MnO_3$ (LCMO) is one of the paradigmatic phase-separated manganites. It presents at room temperature orthorombic structure with *Pnma* space group and cell parameters a = 0.542 nm, b = 0.765 nm and c= 0.543 nm [34], which can also be described by a pseudo-cubic lattice with cell parameter $a_{ps}$ = 0.382 nm ≈ a/√2 ≈ c/√2 ≈ b/2. Early reports in LCMO suggested the existence of a first transition, from a paramagnetic-insulating (PM-I) state to FM-M state at $T_C$ ≈ 225 K, followed by another transition to a AFM-I (also charge ordered) state at $T_N$ ≈ 155 K [35]. However, it was shown later that both phases coexist at temperatures below $T_C$ [36] and that even at temperatures below $T_N$ some FM-M phase persists [37]. It was proposed for epitaxial pulsed lased deposited LCMO thin films, grown on different substrates, that a large strain -either tensile or compressive- favours carrier localization and an insulating behavior while a lower strain -leading to a fairly cubic structure- favours the presence of FM-M phase and triggers the appearance of an insulator-metal transition at ≈ 130-150 K [26].

Besides the potential interest of phase separated LCMO-based devices for neuromorphic computing, only few reports dealing with the memristive properties of LCMO thin films can be found in the literature [29, 30, 38]. In this work we explore in epitaxial LCMO thin films the possibility of obtaining volatile and non-volatile memristive effects, at different temperatures, from a combination of self-heating effects and FM-M phase growth induced by the application of external electric field. Our results show the potential of these systems for obtaining both synaptic and neuron-like behavior in single devices.

Epitaxial LCMO thin films were grown on (100) oriented $SrTiO_3$ (STO) single crystal substrates by pulsed laser deposition from a ceramic stochiometric target. The growth temperature and oxygen pressure were fixed at 648 °C and 0.2 mbar, respectively, while a laser fluence of 0.5 Jcm$^{-2}$ was used. After growth, the sample was annealed in O$_2$ during cool down. X-ray characterization was performed with Panalytical Empyrean diffractometer. High resolution Scanning Transmission Electron Microscopy (STEM) was done using a FEI Titan G2 microscope with a probe corrector (60–300 keV). In-situ chemical analysis was performed by Energy Dispersive Spectroscopy (EDS) and Electron Energy Loss Spectroscopy (EELS). Samples for TEM were prepared by Focused Ion Beam (FIB) in a Helios 650 dual beam equipment. Magnetic properties were measured with a Quantum Design MPMS-XL 7T SQUID magnetometer, between room temperature and 50 K, with the magnetic field applied in the plane of the films. The electrical transport properties were measured between 35K and room temperature, with the Van der Pauw geometry, by using a Keithley 2400 source meter unit hooked to a Lakeshore cryostat. Au electrodes (70 nm thick) were sputtered through a shadow mask.

Fig. 1(a) displays the X-ray characterization performed on a LCMO thin film on STO. A LCMO thickness of ≈ 63 nm was estimated from the X-ray reflectivity (XRR) measurement displayed in the inset. The Bragg-Brentano scan, shown in the main panel, evidences the epitaxial character of the film, characterized by the sole presence of (00h) reflections -we refer to the pseudo-cubic perovskite lattice-. The out-of-plane (OOP) lattice parameter was extracted from the Bragg-Brentano pattern with a value $a_{OOP}$ = 0.376(1) nm, which is 1.6% lower than $a_{pc}$ = 0.382 nm, the value corresponding to the bulk compound. This indicates the presence of in-plane tensile strain, as expected given the positive mismatch between the pseudo-cubic LCMO bulk structure and the cubic STO cell ($a_{STO}$ = 0.3905 nm). Fig. 1(b) displays a high-resolution STEM High-Angle Annular Dark Field (HAADF) cross-section of the same sample. A LCMO thickness of ≈ 66 nm is observed, in good agreement with the one extracted from XRR. The image also evidences the high structural quality of the LCMO film. The STO/LCMO interface is well-defined with no signs of interdifussion, as shown in the inset of Fig. 1(b), where the cations have been labelled. The LCMO layer chemical composition was quantified by EDS and EELS. Fig. 1(d) shows the atomic percentages for the different cations and oxygen as a function of the distance to the LCMO surface, extracted from the EDS line-scan displayed in the low magnification STEM-HAADF image shown in Fig. 1(c). The main feature related to the chemical quantification is the presence of some La excess and Ca deficiency. From the combined EDS/EELS analysis, the average LCMO composition is estimated as $La_{0.56(1)}Ca_{0.44(1)}Mn_{0.96(1)}O_{3.00(1)}$. According to the $La_{1-x}Ca_xMnO_3$ phase diagram [39], a Ca deficiency should favour the FM-M state, which in thin films with tensile strain like ours might counter-back their tendency to present a more robust AFM-I state [26].

Fig. 2(a) shows the temperature-dependent magnetization [M(T)] recorded on a STO/LCMO thin film under field-cooled-cooling (FCC, magnetic field $H$ =0.1 T applied along the in-plane [100] direction). It is seen an increase of M as T decreases, with an upturn around ≈ 250 K, characteristic of a PM to FM transition. In order to determine more precisely the Curie temperature ($T_C$) we have analyzed the temperature evolution of the first derivative of the magnetization, shown in the bottom-left inset of Fig. 2(a). From the position of the minimum, we determine $T_C$ ≈ 225(5) K, in good agreement with the values reported for bulk LCMO [35]. The up-right inset of the same figure shows a M vs. H loop recorded at 150 K, where the presence of hysteresis is evident, with a remanent magnetization $M_R$ ≈ 0.025 $\mu_B$/Mn. From the same loop we extracted a ferromagnetic saturation magnetization $M_S$ ≈ 0.24 $\mu_B$/Mn, in good agreement with previous reports on the same system [30]. This suggests that only a fraction (≈ 7 % at 150 K) of the LCMO film is ferromagnetic and this fraction is embedded in a non-FM matrix. The presented magnetic measurements do not evidence the transition to the AFM state -$T_N$ ≈ 155 K in bulk [35]-, which could be easily masked by the presence of FM-M clusters persisting at low temperatures [37].

In order to get further evidence of the existence of magnetic phase separation and the possibility of changing the balance between FM-M and non-FM phases with an external magnetic field, we performed the experiments shown in Fig. 2(b). The sample was cooled in zero magnetic field from room temperature to 200 K, 140 K and 50 K -all below the extracted $T_C$-, a field $H$ =0.5 T was applied afterwards and the evolution of M with time (t) was recorded for approximately 2 hours. It is found for all temperatures that M increases with t (≈ 13 %, 10 % and 15 % for 200, 140 and 50 K, respectively), reflecting a growth of the FM-M phase at expenses of the non-FM matrix.

We turn now to the transport properties. Fig. 3(a) shows the temperature dependence of the resistance [R(T)], recorded by applying a constant electric field of 14x10$^4$ V/m. Data was taken during warming at a 3 K/min rate. It is seen the presence of a cusp at $T_{MI}$ ≈ 50-60 K [40], indicating a metal-to-insulator transition upon warming. The positive temperature coefficient seen at low temperatures reflects the existence of FM-M phase above the percolation limit (≈ 10-25 %, according to Ref. [32]). The inset shows the presence of a slight hysteresis between cooling and warming measurements, in the range ≈ 60-130 K, which is usually found in phase separated systems [25]. It is also seen a subtle bump at ≈ 130-135 K which might be an indication of the AFM-I transition [30]. Fig. 3(b) shows R(T) curves recorded on cooling under applied electrical fields between $1.6 \times 10^4$ and $16 \times 10^4$ V/m. It is found a large resistance drop, of up to two orders of magnitude, for temperatures close to $\approx 200$ K, when the applied electrical field is increased. This provides a clear indication that the application of electrical field favours the growth of the FM-M phase, an effect that is more pronounced in the temperature range between $T_{MI}$ and $T_C$.

Next, we explored the existence of memristive effects at different temperatures, as shown in Figs. 4(a)-(c). For these experiments, we cooled the sample from room temperature to 50, 140 and 200 K, respectively, in zero electrical field, and then we measured the evolution of R with t upon the application of a constant $14 \times 10^4$ V/m electrical field during 600 s, followed by a period of 200 s with no electrical field applied. We repeated this process for 6-8 times. At 50 K, it is seen that the resistance growths exponentially upon the application of the field (R(600 s)/R(0 s) $\approx$ 1.65), with a time constant of $\approx 13$ s (see Fig. 4(d)). When the electrical field is removed, the resistance returns to a value close to the initial state, indicating the presence of a volatile memristive effect. We notice that a volatile memristive effect implying an increase of the resistance is quite unusual, as pointed out by Salev et al. [41], where they found for $La_{0.7}Sr_{0.3}MnO_3$ a similar effect related to the formation of an insulator barrier, in the direction perpendicular to the current flow, driven by self-heating effects in the vicinity of the metal-insulator transition. In our case, 50 K is close to the metal-insulator transition found in our films and thermal effects are likely behind the observed resistive change -the dissipated power ($P = V^2/R$) at that temperature is $\approx 0.76$ $\mu$W-. At this temperature, FM-M phase enlargement driven by the external electric field does not seem to play a relevant role in the observed resistive changes. The resistance relaxations are quite different at 140 and 200 K. In both cases, it is found that R decreases exponentially after the application of the electrical field. R(600 s)/R(0 s) is $\approx 0.92$ and $\approx 0.95$ for 140 and 200 K, respectively, and the time constants of the relaxations are $\approx 39$ s and $\approx 17$ s, respectively (see Figs. 4(e)-(f)). The resistance drop in these cases is driven by a combination of self-heating (we recall the negative temperature coefficients for the resistance at those temperatures), with dissipated powers estimated in $\approx 71$ $\mu$W and $\approx 486$ $\mu$W, respectively (both higher than in the 50 K case), in synergy with a FM-M enlargement due to the external electrical field. The latter is confirmed by the progressive decrease of the resistance at the beginning of each cycle, indicated with dotted lines in Figs. 4(b)-(c), suggesting that part of the FM-M phase enlarged during the application of the electrical field remains in this state after its removal. We recall that reversible and non- reversible resistive changes were shown for ceramic Fe-doped LCMO under the application of magnetic stimuli [42]; in our case, the stimuli necessary to produce equivalent changes is fully electric, being therefore more simple to achieve and better suited for device miniaturization. The resistance evolution found at 140 K and 200 K in our case, shown in Fig. 4, indicates the coexistance of volatile and non-volatile memristive effects.

An important issue to address is the possible role of oxygen vacancy (OV) electromigration [43] in the observed non-volatile memristive response. To clarify this, we explored the electrical response of LCMO thin films at room temperature (RT), where the system is in its paramagnetic state, it does not present phase separation and OV dynamics is well known to control the memristive response for other manganite-based systems [43–45]. We found no memristive response at RT (see additional data presented at the Supplementary Information), indicating that for this system -LCMO and Au electrode- OV drift-diffusion does not play a relevant role. At lower temperatures OV dynamics should be necessary slower and less relevant than at RT so we can safely assert that the observed non-volatile memristive effects are indeed related to phase separation.

To conclude, our results show that a constant electrical field could be used as an external knob to produce volatile and non-volatile memristive effects in phase separated LCMO thin films. At low temperatures thermal effects dominate against FM-M phase growth, producing a volatile increase of R during the application of the electrical field. At higher temperatures, the electrical field induced growth of the FM-M starts playing a relevant role and we observed a combined volatile and non-volatile memristive effect. We highlight that phase separated manganites add to the short list of materials displaying both volatile and non-volatile resistive changes [20, 22], which can mimic, on the same device, the electrical behavior of neurons and synapses, a feature that might be useful for the development of neuromorphic computing hardware.

We acknowledge support from ANPCyT (PICT2019-02781, PICT2020A-00415) and EU- H2020-RISE projects MELON (Grant No. 872631) and SPICOLOST (Grant No. 734187). We also acknowledge the LMA-Universidad de Zaragoza for offering access to the microscopy instruments. DR thanks U. Lüders and J. Lecourt for the preparation of the LCMO target and M. Rozenberg for helpful discussions.

**AUTHOR DECLARATIONS**

**Conflict of Interest**

The authors have no conflicts to disclose.

**Author Contributions**

Gerardo Ramírez: investigation, data curation, writing – original draft ; Wilson Román Acevedo: investigation, data curation; Miguel Rengifo: investigation, data curation; J. M. Nuñez: investigation; Myriam Aguirre: investigation, formal analysis, funding acquisition; Javier Briático: investigation, funding acquisition; Diego Rubi: conceptualization, methodology, formal analysis, funding acquisition, supervision, writing – review and editing

**Data availability**

The data that support the findings of this study are available from the corresponding author upon reasonable request.

[1] A. Mehonic and A. Kenyon, Nature **604**, 255 (2022).

[2] S. Yu, *Neuro-Inspired Computing Using Resistive Synaptic Devices* (Springer International Publishing, 2017).

[3] A. Serb, A. Corna, R. George, A. Khiat, F. Rocchi, M. Reato, M. Maschietto, C. Mayr, G. Indiveri, S. Vassanelli, and T. Prodromakis, Sci. Rep. **12**, 2590 (2020).

[4] S.-O. Park, H. Jeong, J. Park, J. Bae, and S. Choi, Nat. Commun. **13**, 2888 (2022).

[5] L. Bégon-Lours, M. Halter, F. Puglisi, L. Benatti, D. Falcone, Y. Popoff, D. Dávila, M. Sousa, and B. Offrein, Adv. Electron. Mater. **8**, 2101395 (2022).

[6] T. Mikolajick, M. Park, L. Begon-Lours, and S. Slesazeck, Adv. Mater. , 2206042 (2022).

[7] S. Oh, H. Hwang, and I. Yoo, APL Mater. **7**, 091109 (2019).

[8] S. Sarwat, B. Kersting, T. Moraitis, V. Jonnalagadda, and A. Sebastian, Nat. Nanotechnol. **17**, 507 (2022).

[9] L. Wang, S.-R. Lu, and J. Wen, Nanoscale Res. Lett. **12**, 347 (2017).

[10] M. Xu, X. Mai, J. Lin, W. Zhang, Y. Li, Y. He, H. Tong, X. Hou, P. Zhou, and X. Miao, Adv. Funct. Mater. **30**, 2003419 (2020).

[11] J. Grollier, D. Querlioz, K. Camsari, K. Everschor-Sitte, S. Fukami, and M. Stiles, Nat. Electron. **3**, 360 (2020).

[12] C. Wang, C. Lee, and K. Roy, Sci. Rep. **12**, 8361 (2022).

[13] K. M. Song, J.-S. Jeong, B. Pan, X. Zhang, J. Xia, S. Cha, T.-E. Park, K. Kim, S. Finizio, J. Raabe, J. Chang, Y. Zhou, W. Zhao, W. Kang, H. Ju, and S. Woo, Nat. Electron **3**, 148 (2020).

[14] A. Sawa, Mater. Today **11**, 28 (2008).

[15] D. Ielmini and R. Waser (Eds.), *Resistive Switching from Fundamentals of Nanoionic Redox Processes to Memristive Device Applications* (John Wiley & Sons, 2015).

[16] M. Prezioso, F. Merrikh-Bayat, B. Hoskins, G. Adam, K. Likharev, and D. Strukov, Nature **521**, 61 (2014).

[17] S. Kumar, X. Wang, J. Strachan, Y. Yang, and W. Lu, Nat. Rev. Mater. **7**, 575 (2022).

[18] H. Jiang, F. Belkin, S. Savel'ev, S. Lin, Z. Wang, Y. Li, S. Joshi, R. Midya, C. Li, M. Rao, M. Barnell, Q. Wu, J. Joshua Yang, and Q. Xia, Nat. Commun. **8**, 882 (2017).

[19] W. Chen, H. Barnaby, and M. Kozicki, IEEE Electron Device Lett. **37**, 580 (2016).

[20] E. Janod, J. Tranchant, B. Corraze, M. Querré, P. Stoliar, M. Rozenberg, T. Cren, D. Roditchev, V. Phuoc, M.-P. Besland, and L. Cario, Adv. Funct. Mater. **25**, 6287 (2015).

[21] J. Tian, Z. Tan, Z. Fan, D. Zheng, Y. Wang, Z. Chen, F. Sun, D. Chen, M. Qin, M. Zeng, X. Lu, X. Gao, and J.-M. Liu, Phys. Rev. Appl. **11**, 024058 (2019).

[22] C. Ferreyra, M. Rengifo, M. Sánchez, A. Everhardt, B. Noheda, and D. Rubi, Phys. Rev. Appl. **14**, 044045 (2020).

[23] R. Wang, J.-Q. Yang, J.-Y. Mao, Z.-P. Wang, S. Wu, M. Zhou, T. Chen, Y. Zhou, and S.-T. Han, Adv. Intell. Syst. **2**, 2000055 (2020).

[24] A. Moreo, S. Yunoki, and E. Dagotto, Science **283**, 2034 (1999).

[25] P. Levy, F. Parisi, G. Polla, D. Vega, G. Leyva, H. Lanza, R. Freitas, and L. Ghivelder, Phys. Rev. B **62**, 6437 (2000).

[26] D. Rubi, S. Duhalde, M. Terzzoli, G. Leyva, G. Polla, P. Levy, F. Parisi, and R. Urbano, Phys. B: Condens. Matter **320**, 86 (2002).

[27] M. Uehara and S.-W. Cheong, Europhys Lett. **52**, 674 (2000).

[28] A. Anane, J.-P. Renard, L. Reversat, C. Dupas, P. Veillet, M. Viret, L. Pinsard, and A. Revcolevschi, Phys. Rev. B **59**, 77 (1999).

[29] S. Duhalde, M. Villafuerte, G. Juárez, and S. Heluani, Phys. B: Condens. Matter **354**, 11 (2004).

[30] Z.-H. Wang, Q. Zhang, G. Gregori, G. Cristiani, Y. Yang, L. G. X. Li, J. Sun, B.-G. Shen, and H.-U. Habermeier, Phys. Rev. Mater. **2**, 054412 (2018).

[31] J. Jeon, J. Jung, and K. Chow, J. Phys. D: Appl. Phys **54**, 315002 (2021).

[32] Y. Liu, Z. Xu, K. Qiao, F. Shen, A. Xiao, J. Wang, T. Ma, F. Hu, and B. Shen, Nanoscale **13**, 8030 (2021).

[33] S. Banik, K. Das, K. Pradhan, and I. Das, Europhys. Lett. **133**, 17006 (2021).

[34] P. G. Radaelli, D. E. Cox, M. Marezio, and S.-W. Cheong, Phys. Rev. B **55**, 3015 (1997).

[35] P. Radaelli, D. Cox, M. Marezio, S.-W. Cheong, P. Schiffer, and A. Ramirez, Phys. Rev. Lett. **75**, 4488 (1995).

[36] S. Mori, C. Chen, and S.-W. Cheong, Phys. Rev. Lett. **81**, 3972 (1998).

[37] M. Roy, J. Mitchell, A. Ramirez, and P. Schiffer, Phys. Rev. B **58**, 5185 (1998).

[38] Z.-H. Wang, Y. Yang, L. Gu, H.-U. Habermeier, R.-C. Yu, T.-Y. Zhao, J.-R. Sun, and B.-G. Shen, Nanotechnology **23**, 265202 (2012).

[39] C. Rao, J. Phys. Chem. B **104**, 5877 (2000).

[40] The temperature at which the cusp shows up was find to be dependent on the applied electrical field and on the temperature change rate (to be reported elsewhere).

[41] P. Salev, L. Fratino, D. Sasaki, R. Berkoun, J. del Valle, Y. Kalcheim, Y. Takamura, M. Rozenberg, and I. Schuller, Nat. Commun. **12**, 5499 (2021).

[42] P. Levy, F. Parisi, L. Granja, E. Indelicato, and G. Polla, Phys. Rev. Lett. **89**, 137001 (2002).

[43] M. J. Rozenberg, M. J. Sánchez, R. Weht, C. Acha, F. Gomez-Marlasca, and P. Levy, Phys. Rev. B **81**, 115101 (2010).

[44] W. Román Acevedo, C. Acha, M. J. Sánchez, P. Levy, and D. Rubi, Appl. Phys. Lett. **110**, 053501 (2017).

[45] W. Román Acevedo, C. A. M. van den Bosch, M. H. Aguirre, C. Acha, A. Cavallaro, C. Ferreyra, M. J. Sánchez, L. Patrone, A. Aguadero, and D. Rubi, Appl. Phys. Lett. **116**, 063502 (2020).

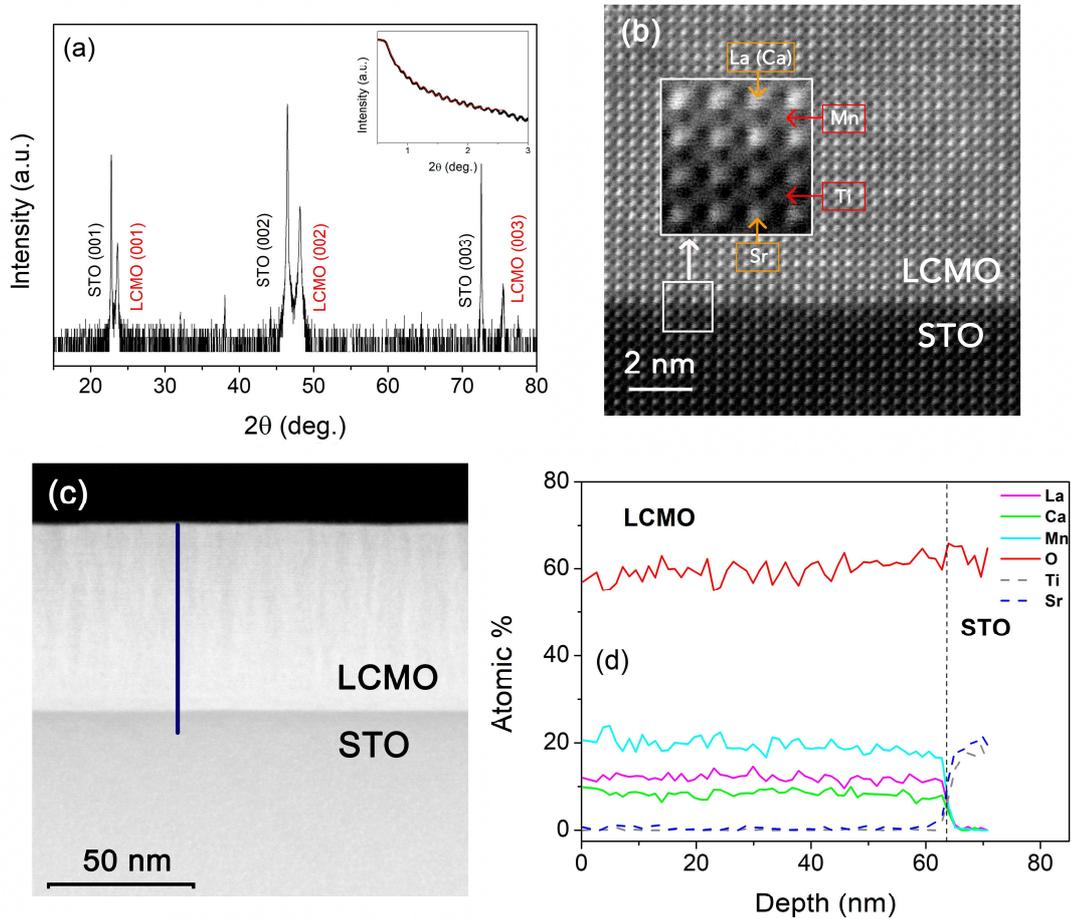

Figure 1: (a) Bragg-Brentano XRD scan corresponding to a LCMO thin film on STO. The inset shows the X-ray reflectivity recorded for the same sample; (b) STEM-HAADF cross section corresponding to the same film. The inset shows a blow-up of the interface, where cations have been labelled; (c) Low magnification STEM-HAADF cross section of the same sample. An EDS linescan was performed in the zone indicated with a blue line. The scan starts in the LCMO layer and ends in the STO substrate; (d) Chemical quantification for cations and oxygen, obtained from the EDS linescan described in (c).

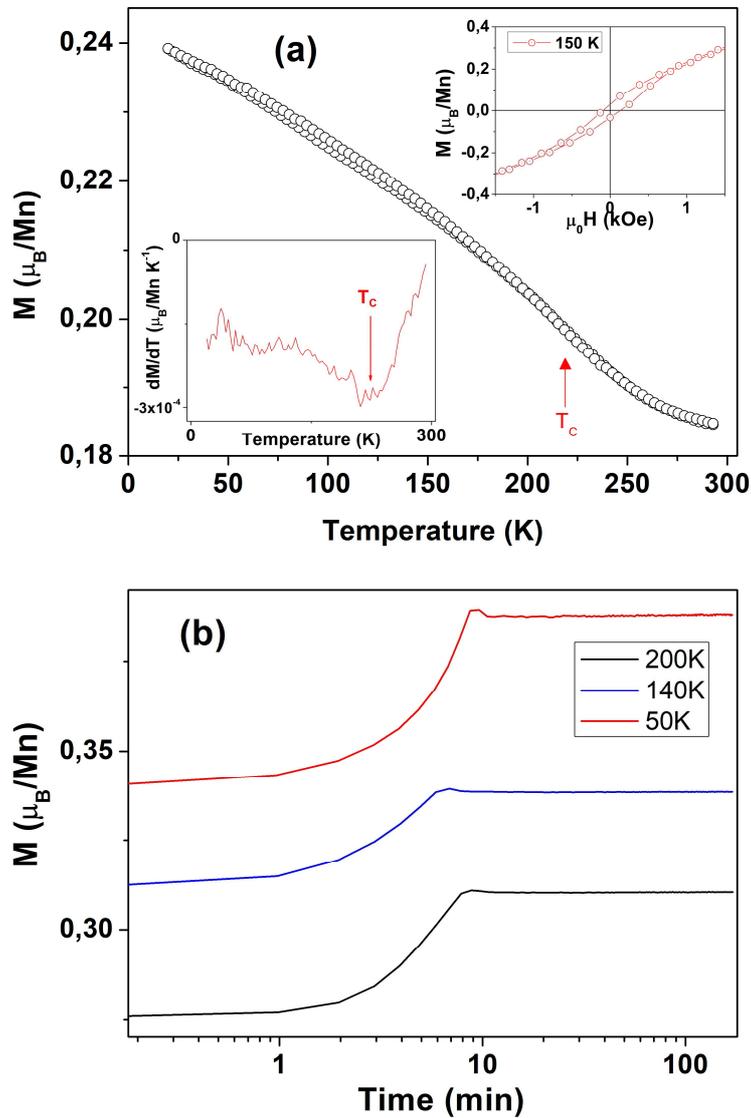

Figure 2: (a) Field-cooled-cooling (FCC) magnetization vs. temperature curve measured on a STO/LCMO film. A field $H$ =0.1 T was applied along the in-plane perovskite [100] direction. The bottom-left inset shows the derivative of M(T) from which the $T_C$ value was extracted. The top-right inset shows a hysteresis loop recorded at 150 K; (b) Magnetic time relaxation curves measured on the same film at 200 K, 140 K and 50 K, with an applied in-plane magnetic field of 0.5 T. See the main text for the details of the measuring procedure

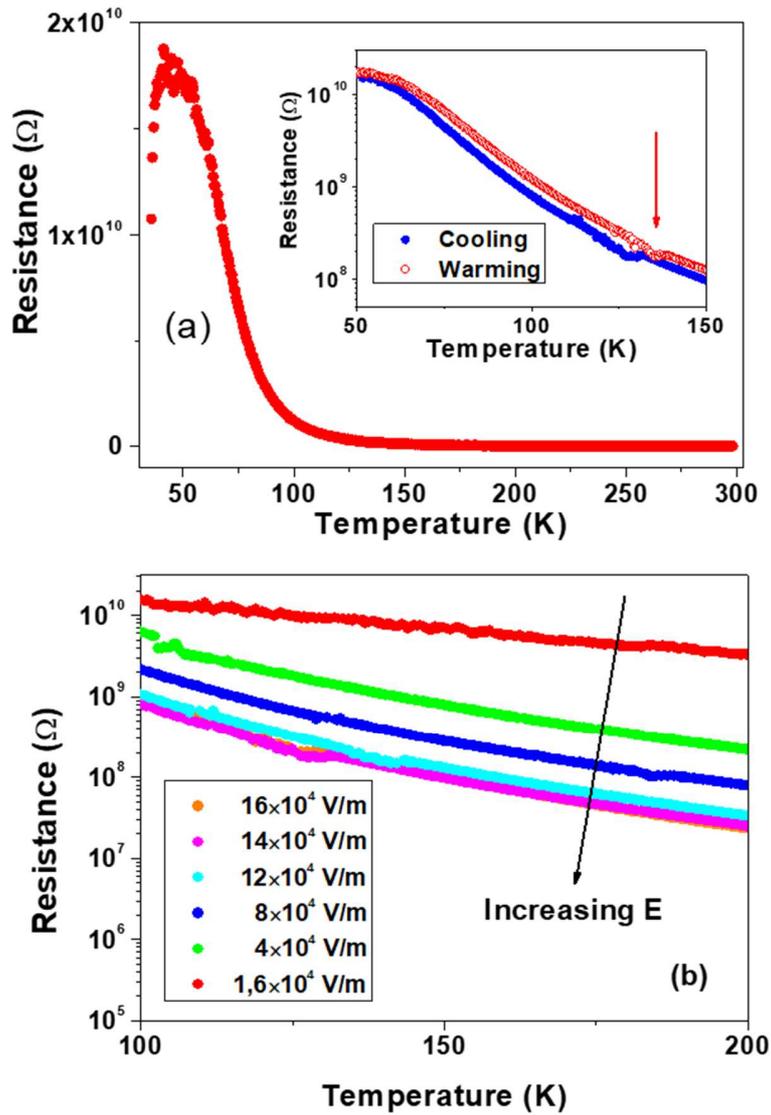

Figure 3: (a) Resistance as a function of the temperature for a STO/LCMO thin film. A constant $14 \times 10^4$ V/m electrical field was applied during the measurement. Data was taken during warming. The inset shows a slight hysteresis observed between warming and cooling measurements. The red arrows indicates the presence of a bump which we attribute to the AFM-I transition; (b) Resistance as a function of the temperature for the same sample, measured on cooling under different applied electrical fields.

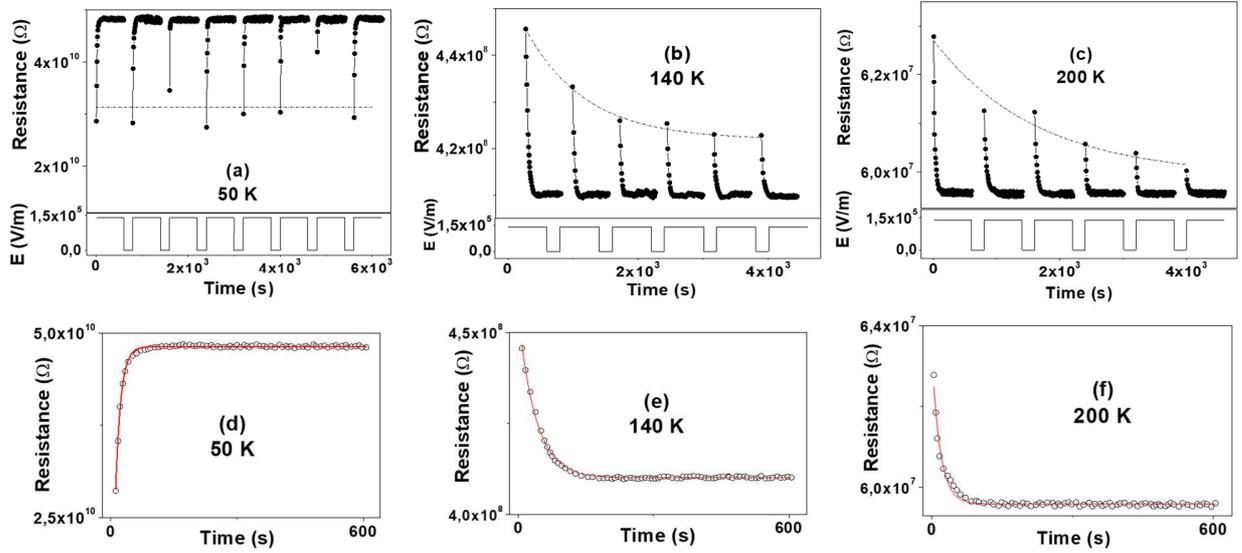

Figure 4: Resistance time relaxations measured on a STO/LCMO sample upon the application of a $14 \times 10^4$ V/m electrical field, during 600 s and for several cycles, at (a) 50 K, (b) 140 K and (c) 200 K. See the main text for details about the measurement procedure. The dashed lines indicate the trend followed by the resistive states at the beginning of each cycle starting with the application of the electrical field; (d), (e), (f) Exponential fits of the first cycle of the resistance relaxations at 50 K, 140 K and 200 K, respectively. The fittings are shown as red solid lines.